\documentclass[showpacs,prb,superscriptaddress,twocolumn]{revtex4}

\usepackage{textcomp}
\usepackage{amsmath}
\usepackage{amssymb}
\usepackage{graphicx} 
\usepackage[lofdepth,lotdepth,caption=false]{subfig}
\usepackage[breaklinks=true,colorlinks,citecolor=blue,linkcolor=blue,urlcolor=blue]{hyperref}

\usepackage{bm}
\usepackage{color}
\usepackage{ulem}
\usepackage{textcomp}

\newcommand{\e}{e}

\newcommand{\us}{\uparrow}
\newcommand{\ds}{\downarrow}

\begin{document}

\title{Thermopower of three-terminal topological superconducting systems}

\author{Stefano Valentini}
\affiliation{NEST, Scuola Normale Superiore and Istituto Nanoscienze-CNR, I-56127 Pisa, Italy}
\email{stefano.valentini@sns.it}

\author{Rosario Fazio}
\affiliation{NEST, Scuola Normale Superiore and Istituto Nanoscienze-CNR, I-56127 Pisa, Italy}

\author{Vittorio Giovannetti}
\affiliation{NEST, Scuola Normale Superiore and Istituto Nanoscienze-CNR, I-56127 Pisa, Italy}

\author{Fabio Taddei}
\affiliation{NEST, Istituto Nanoscienze-CNR and Scuola Normale Superiore, I-56126 Pisa, Italy}

\begin{abstract}
We study the thermopower of a three-terminal setup composed of a quantum dot attached to three electrodes, one of which is a topological superconductor. In the model, superconductivity is explicitly taken into account. We compare the results for $s$-wave (trivial) and $p$-wave (topological) superconductors and observe that for small temperatures the thermopower has different sign in the two cases. This behavior is strongly dependent on temperature and we estimate an energy scale that controls the sign in the $p$-wave case, which results proportional to the square root of the gap and the coupling to superconductor. The analytical results obtained with a simple 1D model are confirmed by a more realistic tight-binding model.
\end{abstract}


%73.23.-b		Electronic transport in mesoscopic systems
%74.45.+c	Proximity effects; Andreev reflection; SN and SNS junctions
%72.20.Pa 	Thermoelectric and thermomagnetic effects

%74.25.fg		Thermoelectric effects [in 74. Superconductivity]
%74.78.Na	Mesoscopic and nanoscale systems

\maketitle

%%%%%%%%%%%%%%%%%%%%%
\section{Introduction}

In the last few years, a great deal of effort has been put on the study of topological superconducting systems \cite{alicea2012,beenakker2013,leijnse2012} because of the possibility of manipulating quantum information in a protected way \cite{nayak2008}. A possible implementation consists in the realization of an effective (1D) $p$-wave superconductor \cite{kitaev2001}.
This makes use of a semiconducting wire in the presence of spin-orbit interaction, Zeeman fields, and a superconducting order parameter induced by an $s$-wave superconductor located in proximity of the nanowire \cite{lutchyn2010,oreg2010}. In a certain range of parameters, the nanowire is predicted to be in a topologically non-trivial phase, exhibiting a pair of Majorana bound states (MBS) at its ends. 
There has been a great deal of work devoted in understanding how MBS could be detected. Smoking gun proofs of the existence of MBS are, for example,  the $2e^2/h$ quantization of the  
conductance~\cite{sengupta2001,law2009,flensberg2010,wimmer2011} or the fractional Josephson effect with 4$\pi$ periodicity~\cite{kitaev2001,kwon2004}.
The experiments performed so far \cite{mourik2012,das2012,deng2012,finck2013,churchill2013} have measured the differential conductance peak at zero voltage in a two-terminal device.

Recently, the opportunity of using a three-terminal setup has been put forward in Refs.~\onlinecite{valentini2014,weithofer2014} in order to reveal the peculiar behavior of Andreev bound states, which form between two topological superconducting nanowires when a phase bias is applied.
In Ref.~\onlinecite{valentini2014}, in particular, additional peculiar features in the spectrum of Andreev levels, due to the presence of a third electrode, have been singled out.

It is also possible to look for effects of topological nature in thermoelectrical properties.
Recently, the thermopower of a topological system has been studied in two-terminal geometries \cite{leijnse2014,lopez2014}.
While Ref. \onlinecite{leijnse2014} focuses on the case of a quantum dot (QD) coupled to a normal lead and to an MBS, Ref. \onlinecite{lopez2014} studies the case of two normal leads coupled to a QD that is side-coupled to an MBS. In the first case the thermal bias is applied between the normal lead and the QD-MBS block, while in Ref. \onlinecite{lopez2014}  the thermal bias is applied between the normal leads. In both cases the presence of a topological superconductor is not explicit but is taken into account within a low-energy effective theory.

In this work we study the thermopower of a three-terminal hybrid system focusing on the differences between trivial and topological phases. We analyze both analytically and numerically a setup composed of a superconducting nanowire attached to two normal leads (see Fig.~\ref{fig:setup_simple}) and we explore different regimes by varying the parameters of the system. We first consider a simplified continuous 1D  model with a single-level QD and a $p$-wave (or $s$-wave) superconducting wire. 
We find that the sign of the thermopower for the $p$ wave is reversed with respect to the $s$ wave for small temperatures. This behavior is strongly temperature-dependent and disappears for higher temperatures.
We find that the energy scale that controls the transition between those behaviors is $\sqrt{\Delta \Gamma}$, where $\Delta$ is the induced superconducting gap and $\Gamma$ is the coupling to the superconducting lead.
We then consider a more realistic description of the system using a 2D tight-binding model of a nanowire in the presence of Zeeman and spin-orbit effects, and $s$-wave superconducting coupling. By tuning the parameters one can now explore intermediate regimes between $s$-wave- and $p$-wave-like superconductivity including the topological phase transition. Numerical simulations confirm the results of the simple model and extend them to more general scenarios such as many-level QD and many-channel normal leads. The behavior across the topological phase transition is also analyzed.
We complete our study by considering a topological system where one of the two normal leads is removed (two-terminal setup), and nontopological systems in the presence of zero-energy fermionic levels.

The paper is organized as follows. In Sec.~\ref{system}, we describe the system and the framework we use, and in Sec.~\ref{1d_model}, we analyze in details a simple 1D continuous model and present the results on the thermopower in Sec. \ref{results}. In Sec.~\ref{tightbind}, we study a more realistic 2D tight-binding model, while in the Appendixes we analyze two-terminal or nontopological related systems.

\section{System and framework}\label{system}
\begin{figure}[tb]
\centering
\includegraphics[width=0.9\columnwidth, keepaspectratio]{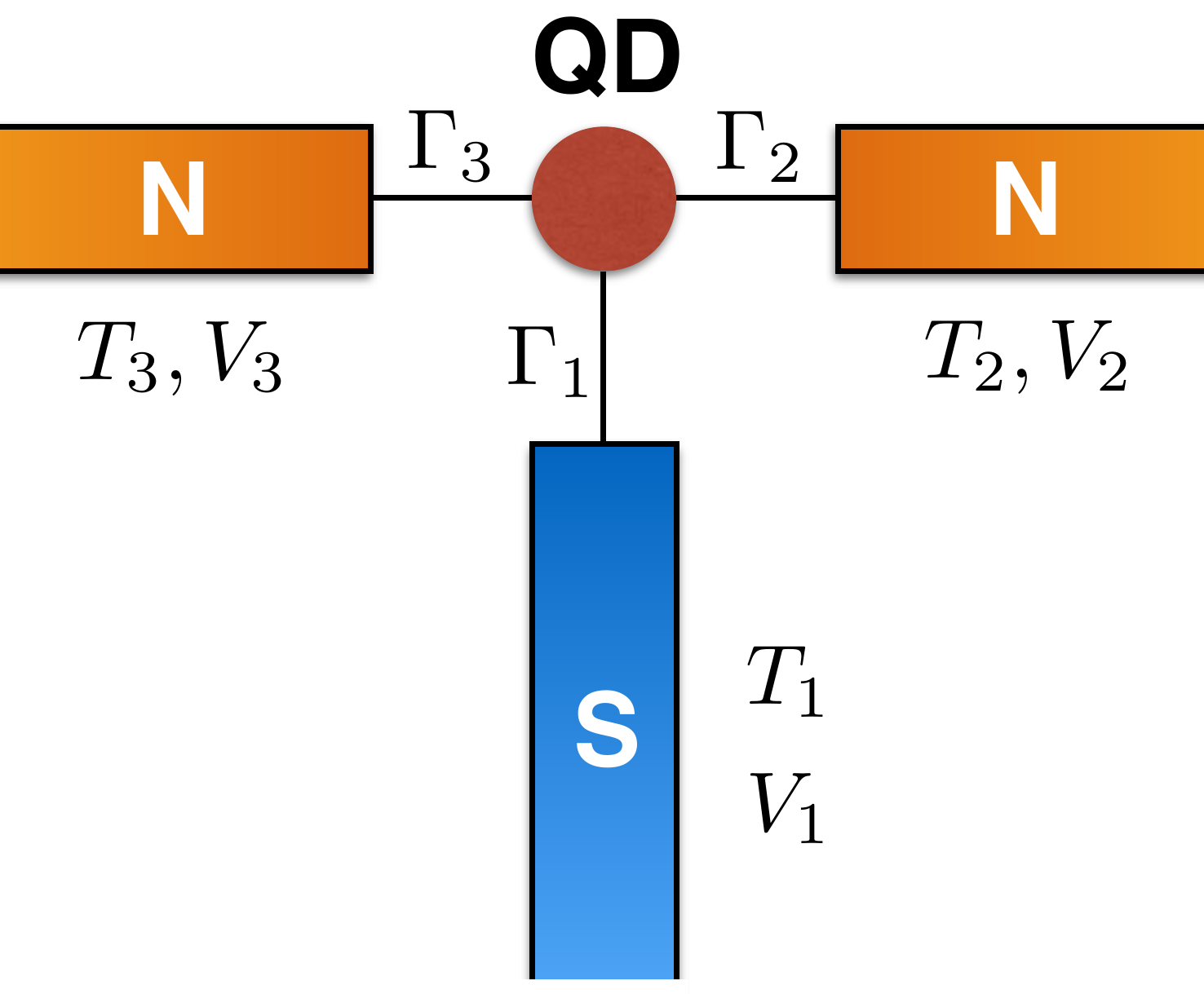}
\caption{Scheme of the setup: QD coupled to two normal leads (N) and a topological superconducting wire (S).}
\label{fig:setup_simple}
\end{figure}

The system under investigation is schematically drawn in Fig.~\ref{fig:setup_simple}. A QD is coupled to two normal leads (terminals 2 and 3) and to a superconducting wire (terminal 1) that can be either topological or not. Each lead is characterized by a temperature $T_i$, a voltage $V_i$ and a coupling to the QD $\Gamma_i$, with $i=1,2,3$. We choose the superconductor (lead 1) as the reference and define the following voltage and temperature biases: $\Delta V_2=V_2 - V_1$, $\Delta V_3=V_3 - V_1$,  $\Delta T_2=T_2 - T_1$, $\Delta T_3=T_3 - T_1$.

In the linear-response regime, the relation between the heat/charge currents and the applied biases reads
\begin{equation}
\begin{pmatrix}
J_N^2 \\
J_Q^2 \\
J_N^3\\
J_Q^3 \\
\end{pmatrix}
= 
\begin{pmatrix}
\mathcal{G}_{22} & \mathcal{D}_{22} & \mathcal{G}_{23} & \mathcal{D}_{23} \\
\mathcal{M}_{22} & \mathcal{K}_{22} &\mathcal{M}_{23} & \mathcal{K}_{23} \\
\mathcal{G}_{32} & \mathcal{D}_{32} & \mathcal{G}_{33} & \mathcal{D}_{33} \\
\mathcal{M}_{32} & \mathcal{K}_{32} &\mathcal{M}_{33} & \mathcal{K}_{33} \\
\end{pmatrix}
\begin{pmatrix}
\Delta V_2 \\
\Delta T_2 \\
\Delta V_3 \\
\Delta T_3
\end{pmatrix},
\end{equation}
where $J_N^i (J_Q^i)$ is the charge (heat) current for lead $i=2,3$ and $\mathcal{G}_{ij}$, $\mathcal{D}_{ij}$, $\mathcal{M}_{ij}$, $\mathcal{K}_{ij}$ are the Onsager coefficients.

The thermopower in a multiterminal system can be defined as in Ref. \onlinecite{mazza2014}, 
where the local thermopower relative to lead 2 is
\begin{equation} \label{eq::thermo}
 \left.\mathcal{S}=-\frac{\Delta V_2}{\Delta T_2} \right|_{J_N^1=0,J_N^2=0,J_N^3=0,T_3=T_1},
\end{equation}
which can be rewritten in terms of Onsager coefficients as follows:
\begin{equation}
\label{thermopower}
\mathcal{S}=\frac{\mathcal{D}_{22} \mathcal{G}_{33}-\mathcal{D}_{32} \mathcal{G}_{23}}{\mathcal{G}_{22}\mathcal{G}_{33}-\mathcal{G}_{32}\mathcal{G}_{23}}.
\end{equation}
Using the Landauer-B\"{u}ttiker theory \cite{landauer1957, buttiker1986, claughton1996} 
one can express the Onsager coefficients in terms of the scattering probabilities between leads:
\begin{equation}
\label{G}
\begin{aligned}
  \mathcal{G}_{ij} &= \frac{e^2}{h} \int_0^{+\infty}dE \, \Big[ \sum_{\alpha\sigma}N_i^{\alpha\sigma}(E) \delta_{ij} \\
  &-  \sum_{\alpha\sigma\beta\sigma'}   \alpha\beta \, P_{ij}^{\alpha\sigma\beta\sigma'}(E)\Big] \Big(-\frac{\partial f(E)}{\partial E}\Big) ,
\end{aligned}
\end{equation}
\begin{equation}
\label{D}
\begin{aligned}
 \mathcal{D}_{ij} &= \frac{e}{h} \int_0^{+\infty} d E \, \frac{E}{T} \Big[ \sum_{\alpha\sigma} \delta_{ij} \alpha  N_i^{\alpha\sigma}(E)  \\ 
&- \sum_{\alpha\sigma\beta\sigma'}  \alpha \, P_{ii}^{\alpha\sigma\beta\sigma'}(E)\Big] \Big(-\frac{\partial f(E)}{\partial E}\Big),
\end{aligned}
\end{equation}
where $\alpha$ and $\beta$ are equal to $+1$ for particles and $-1$ for holes, $e$ is the modulus of the electron charge and $h$ is the Planck constant.
In Eqs.~(\ref{G}) and (\ref{D}), $P_{ij}^{\alpha\sigma\beta\sigma'}(E)$ is the probability for a particle of energy $E$ of type $\beta$ and spin $\sigma'$ from lead $j$ to be scattered as a particle of type $\alpha$ and spin $\sigma$ into lead $i$. $N_i^{\alpha\sigma}(E)$ is the number of open channels at energy $E$ in lead $i$. Note that the energy is measured from the electrochemical potential of the condensate of the superconducting lead and  $f(E)=(1+e^{\frac{E}{k_B T}})^{-1}$ is the Fermi-Dirac distribution with $k_B$ the Boltzmann constant and $T$ the reference temperature 
(we dropped the subscript $1$ relative to the superconducting wire for simplicity, $T \equiv T_1$).

\section{1D continuous model}\label{1d_model}
In order to develop an understanding of the system, let us start analyzing a limit amenable of an analytic solution by
considering a single-level QD coupled to one-dimensional leads. Two leads are normal and one is superconducting, either $p$ wave or $s$ wave.
In order to correctly describe the superconductivity we work in the Nambu basis. 
As far as the $p$-wave case is concerned, we are interested in the spin-less case and we describe the normal leads as 1D spin-less free electron gases.
For the $s$-wave case, the singlet superconducting coupling involves particles and holes of different spin. Since no spin-mixing mechanism exists and no magnetic fields are present, the sector of the Hamiltonian relative to spin-up electron and spin-down hole is decoupled from (and degenerate with) the sector of spin-down electron and spin-up hole, hence we focus on only one of these two sectors. Since the spin for both electrons and holes is fixed, we drop any reference to the spin of the particles and describe the system as for the $p$-wave case except for the $s$-wave superconducting coupling. 

If the three leads are normal, the scattering matrix, in the wide-band approximation, is given by \cite{buttiker1988}
\begin{equation}\label{qdmatrix}
S^{\text{QD},ee}_{ij}(E) = \delta_{ij} - \frac{ i \sqrt{\Gamma_i \Gamma_j}}{ (E-E_D) + i \frac{\Gamma}{2} },
\end{equation}
where $E_D$ is the QD level, $\Gamma_i$ is the coupling to the $i$-th lead and $\Gamma=\Gamma_1+\Gamma_2+\Gamma_3$. 
In order to calculate the scattering matrix of the whole system we use the particle-hole symmetry relations and compose the scattering matrix of Eq. (\ref{qdmatrix}) with that of a clean N-S junction [any additional barrier at the interfaces between QD and leads would only renormalize the parameters of Eq. (\ref{qdmatrix})]. 
The reflection sector of the N-S scattering matrix expressed in the electron-hole basis is 
\begin{equation}
R_{NS}(E) = 
\begin{pmatrix}
0 & \e^{- \mathrm{arccosh}{(E/\Delta)} } \\
 \e^{- \mathrm{arccosh}{(E/\Delta)} } & 0
\end{pmatrix},
\end{equation}
for the $s$-wave case, and
\begin{equation}
R_{NS}(E) = 
\begin{pmatrix}
0 & \e^{- \mathrm{arccosh}{(E/\Delta)} } \\
- \e^{- \mathrm{arccosh}{(E/\Delta)} } & 0
\end{pmatrix},
\end{equation}
for the $p$-wave case. The inverse hyperbolic cosine has to be intended with its analytic continuation and $\Delta$ is the superconducting gap. 
By composing the scattering matrices defined above, one can find the transmission and reflection coefficients and, by means of 
 Eqs.~(\ref{G}) and (\ref{D}), determine the Onsager matrix. The transmission sector is not needed in the following and is null for energies inside the gap. 
 The results are discussed in the following.
 
\subsection{Results}\label{results}

\begin{figure}[tb]
\centering
\includegraphics[width=0.9\columnwidth, keepaspectratio]{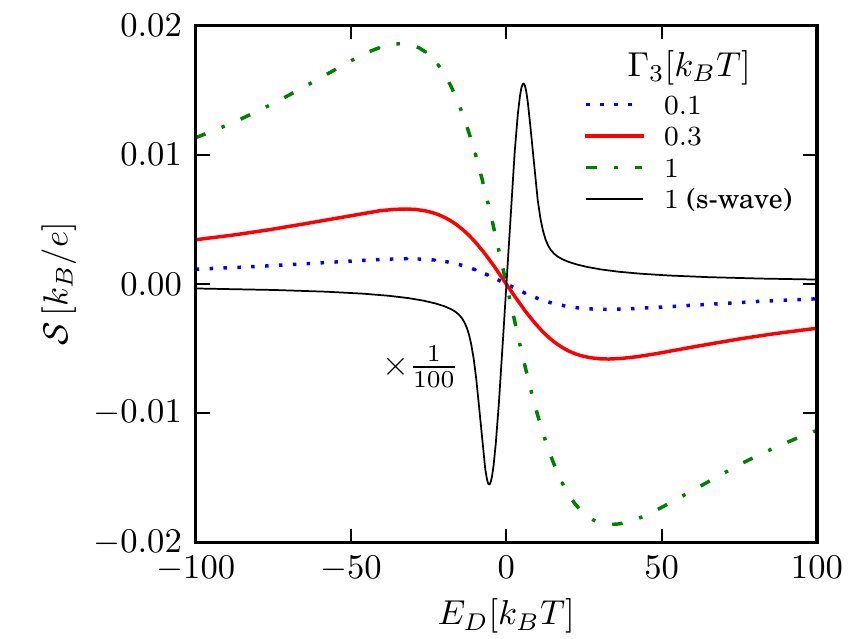}
\caption{$\mathcal{S}$ of the 1D continuous model as a function of $E_D$ for $\Delta=100 k_B T$, $\Gamma_1=10 k_B T$, $\Gamma_2=0.1 k_B T$, and different values of $\Gamma_3$ for the $p$-wave case, and $s$-wave case (divided by a factor 100) for comparison.}
\label{fig:g3}
\end{figure}

In Fig. \ref{fig:g3}, the thermopower $\mathcal{S}$  is plotted for the $p$-wave case as a function of the QD level $E_D$ for one choice of the couplings $\Gamma_1$ and $\Gamma_2$ and for different values of $\Gamma_3$. We observe that $\mathcal{S}$ is anti-symmetric in $E_D$, it is linear around $E_D=0$ and shows an extremum, whose amplitude increases with $\Gamma_3$.

\begin{figure}[tb]
\centering
\subfloat[][]{\label{g1}
\includegraphics[width=0.8\columnwidth, keepaspectratio]{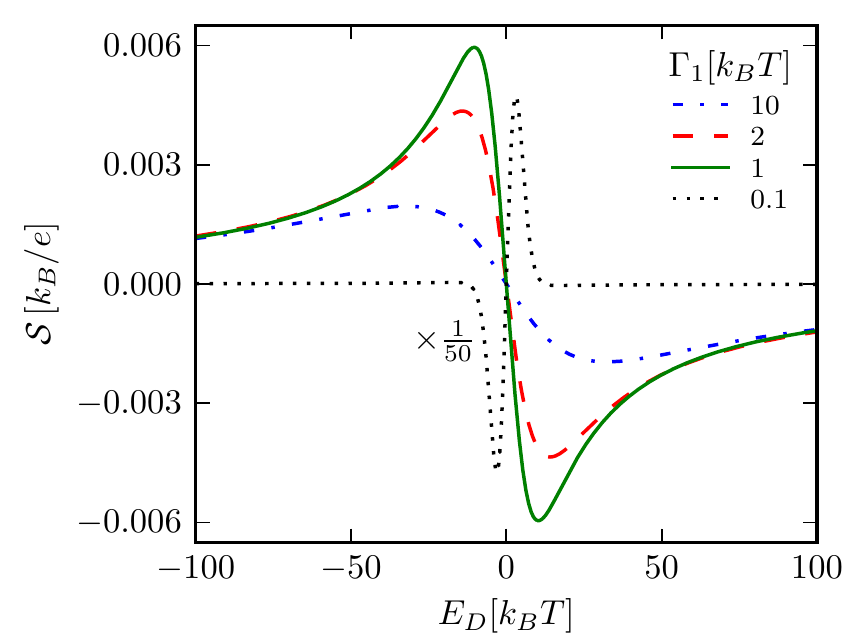}} \\
\subfloat[][]{\label{g1medium}
\includegraphics[width=0.8\columnwidth, keepaspectratio]{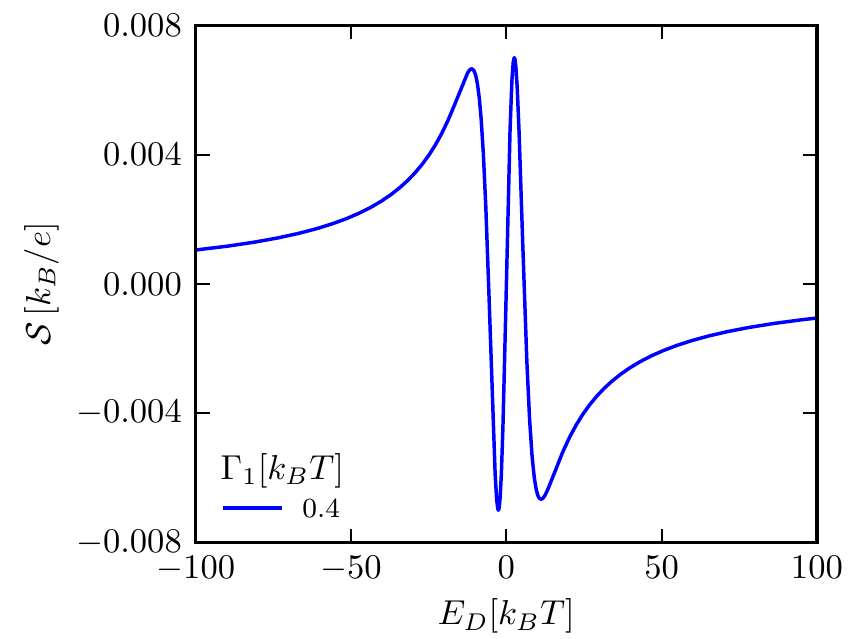}}\llap{\raisebox{0.35 \columnwidth}{\includegraphics[width=0.3\columnwidth]{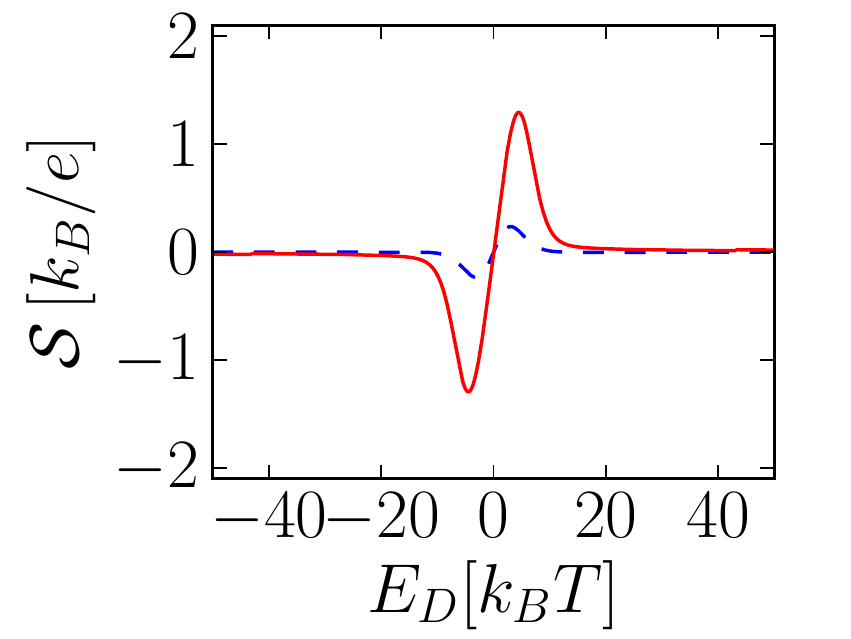}}}
\caption{$\mathcal{S}$ of the 1D continuous model as a function of $E_D$ for $\Delta=100 k_B T$, $\Gamma_2=0.1 k_B T$, $\Gamma_3=0.1 k_B T$ and different values of $\Gamma_1$ for the $p$-wave case. In the inset of (b) the blue dashed line is for $\Gamma_1=0.1 k_B T$ and the red solid line is for $\Gamma_1=0.01 k_B T.$} 
\label{fig:g1}
\end{figure}

\begin{figure}[tb]
\centering
\label{g1sw}
\includegraphics[width=0.9\columnwidth, keepaspectratio]{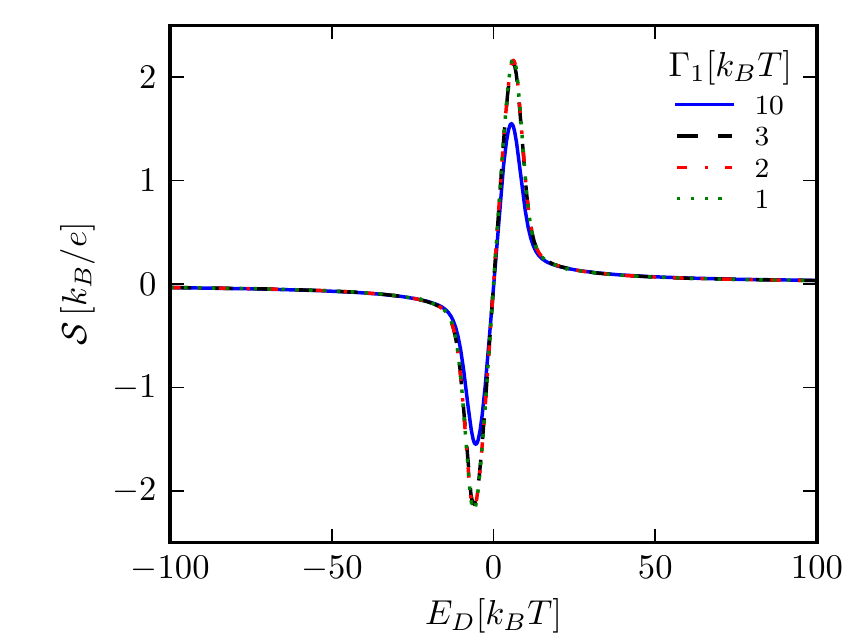}
\caption{$\mathcal{S}$ as a function of $E_D$ for $\Delta=100 k_B T$, $\Gamma_2=0.1 k_B T$, $\Gamma_3=0.1 k_B T$ and different values of $\Gamma_1$ for an $s$-wave superconductor.}
\label{fig:g1sw}
\end{figure}

Most strikingly, the sign of $\mathcal{S}$ is opposite with respect to $s$-wave case (thin black curve in Fig. \ref{fig:g3}), which represents the usual behavior in which the sign of the thermopower reflects the sign of the dominant charge carriers \cite{ashcroft1976}.
This result is in accordance with the findings of the low-energy effective theory of Ref.~\onlinecite{lopez2014}.
For completeness, we checked that the thermopower $\mathcal{S}$ is virtually independent of $\Gamma_2$ in the range between $0.01 k_B T$ and $k_B T$. 
As we shall see in the following, the sign of $\mathcal{S}$ for a $p$-wave superconductor is controlled by the coupling $\Gamma_1$ between QD and superconductor.
As shown in Fig. \ref{fig:g1}, by decreasing $\Gamma_1$ from $10 k_B T$ to $k_B T$ the amplitude of $\mathcal{S}$ increases, eventually changing its sign for $\Gamma_1=0.1 k_B T$. By further decreasing $\Gamma_1$ the thermopower increases keeping the usual sign [see inset of Fig. \ref{g1medium}]. Interestingly, for intermediate values of $\Gamma_1$ a richer structure appears whereby the sign of $\mathcal{S}$ changes three times in the vicinity of $E_D=0$.
We remark that the behavior of the thermopower for the $s$-wave case is trivial when varying $\Gamma_1$. Namely, as shown in Fig. \ref{fig:g1sw}, $\mathcal{S}$ is virtually independent of $\Gamma_1$ for $\Gamma_1\le 3 k_B T$.
We notice that the behavior for vanishingly small $\Gamma_1$ is the same as the one for a QD connected to two normal leads. 

In order to understand the behavior of the thermopower in the $p$-wave case, it is convenient to consider the limit $k_BT\ll\Delta$ and $\Gamma_2 \sim \Gamma_3$.
In this case, the expression of Eq. (\ref{thermopower}) reduces to 
\begin{equation}\label{s_reduced}
\mathcal{S}\simeq \frac{\mathcal{D}_{22}}{\mathcal{G}_{22}},
\end{equation}
 which has the same structure as the two-terminal thermopower (we also verified numerically this approximation in the range of parameters used). From Eqs. (\ref{G}) and (\ref{D}) one has
\begin{equation}
 \begin{aligned}
 \mathcal{G}_{22} &= \frac{e^2}{h} \int_{-\infty}^{+\infty}dE \, G_{22}(E) \Big(-\frac{\partial f(E)}{\partial E}\Big),\\
\mathcal{D}_{22} &= \frac{e}{h} \int_{-\infty}^{+\infty}dE \, \frac{E}{T} G_{22}(E) \Big(-\frac{\partial f(E)}{\partial E}\Big) =\\
&=  \frac{e}{h} \int_{0}^{+\infty}dE \, \frac{E}{T} G_{odd,22}(E) \Big(-\frac{\partial f(E)}{\partial E}\Big),
 \end{aligned}\label{eq:pwave}
\end{equation}
where $G_{22}(E)=1-R^N_{22}(E)+R^A_{22}(E)$ is the zero-temperature local conductance at lead 2, $R_{22}^N=P_{22}^{++}$ is the normal reflection probability at lead 2, $R_{22}^A=P_{22}^{-+}$ is the Andreev reflection probability at the same lead and $G_{odd,22}(E)=G_{22}(E)-G_{22}(-E)$. Eq.~(\ref{eq:pwave}) implies that $G_{22}(E)$ in a range of energy within a few $k_B T$ around zero energy controls the thermopower. 
\begin{figure}[tb]
\centering
\includegraphics[width=0.9\columnwidth, keepaspectratio]{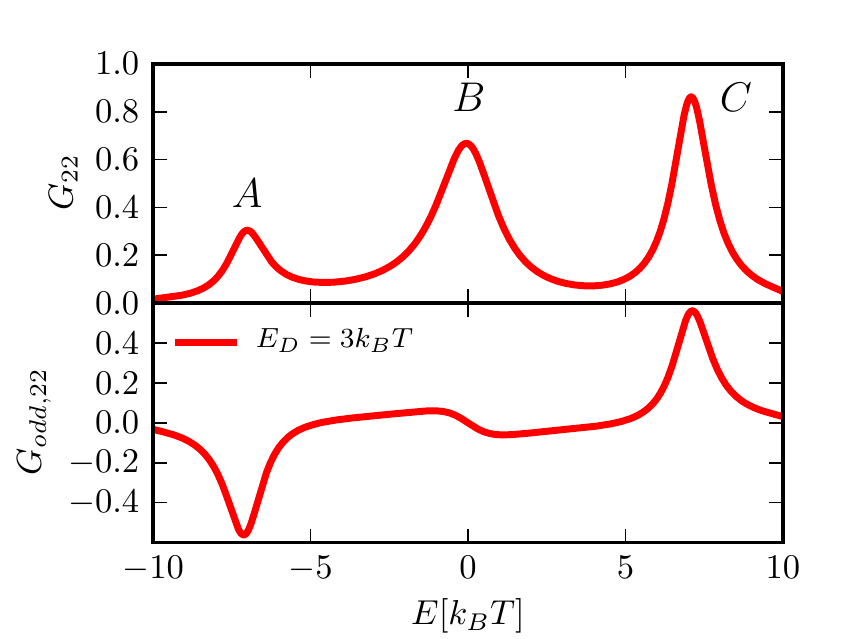}
\caption{ $G_{22}(E)$ and $G_{odd,22}(E)$ as a function of energy $E$ for $\Delta=10 k_B T$, $E_D=3 k_B T$, $\Gamma_1=5 k_B T$, $\Gamma_2=1 k_B T$ and $\Gamma_3=2 k_B T$.}
\label{fig:kernel}
\end{figure}

For very small temperatures, one can use the Sommerfeld expansion, which allows us to obtain 
the Mott formula \cite{mott1958,jonson1980,sivan1986} for the thermopower:
\begin{equation}\label{mott}
\mathcal{S}= \frac{\pi^2 k_B^2 T}{3 e}\Big(\frac{d \ln G_{22}(E)}{dE}\Big)_{E=0}.
\end{equation} 
$G_{22}(E)$ is always positive and Eq. (\ref{mott}) ensures that $\mathcal{S}$ has the same sign as the energy derivative of $G_{22}(E)$ at $E=0$. 
$G_{22}(E)$, plotted in Fig. \ref{fig:kernel} (upper panel), shows a three-peak structure whose symmetry with respect to $E=0$ is broken when $E_D\ne0$. While the two external peaks A and C are related to the QD level $E_D$, the central peak B is due to the presence of the MBS (thorough analysis of the conductance can be found in Refs. \onlinecite{asano2007,liu2011,cao2012,lee2013,vernek2014,gong2014,liu2014}).
The external peak on the same side of $E_D$ (peak C) is higher than the one on the other side (peak A), while the central one (peak B) has the maximum close to zero energy but pushed a little away from the bigger peak C in such a way that, for positive $E_D$, the derivative of $G_{22}(E)$ at zero energy is negative [this is witnessed by the value of $G_{odd,22}$ for positive $E$ close to $E=0$, see Fig.~\ref{fig:kernel} (lower panel)]. 
This makes clear that the Majorana central peak B is responsible for the sign of the thermopower in the small temperature limit.

If the temperature is higher, the Mott formula is no longer applicable and one must calculate the integrals of Eq. (\ref{eq:pwave}). Let us assume, for simplicity, $E_D > 0$. Since $\mathcal{G}_{22}$ is always positive, according to Eq.~(\ref{s_reduced}) the sign of $\mathcal{S}$ is controlled by $\mathcal{D}_{22}$, which depends on the behavior of $G_{odd,22}(E)$ in the interval of energies between 0 and a few $k_B T$. If this interval is large enough (i.e. for high temperatures) such that the integral is dominated by the positive contribution of peak C to $G_{odd,22}$ [see Fig.~\ref{fig:kernel} (lower panel)], then the thermopower will be positive. If the interval is small (i.e. for low temperatures) such that the contribution by peak C is negligible and the negative contribution of peak B dominates the integral, one obtains a negative thermopower recovering the result from the Mott formula [Eq. (\ref{mott})].
More quantitatively, the odd part of $G_{22}(E)$ takes the form: 
\begin{widetext}
\begin{equation}
\label{oddpart}
G_{odd,22}(E)=\frac{\frac{-4 \Gamma_2 \Gamma_3 E E_D}{\Delta \Gamma_1}(\sqrt{1-\frac{E^2}{\Delta^2}}-\frac{2 E^2}{\Delta \Gamma_1})}{\Big[\frac{2E^2}{\Delta^2 \Gamma_1}(E_D^2-E^2+\frac{\Gamma^2}{4})+2\Gamma \frac{E^2}{\Delta \Gamma_1}\sqrt{1-\frac{E^2}{\Delta^2}}-\Gamma_2-\Gamma_3\Big]^2+4E^2\Big[1+\frac{E_D^2-E^2+\frac{\Gamma^2}{4}}{\Delta \Gamma_1}\sqrt{1-\frac{E^2}{\Delta^2}}-\frac{\Gamma E^2}{\Delta^2 \Gamma_1}\Big]^2}.
\end{equation}
\end{widetext}
Assuming $k_B T\ll \Delta$ and very large $|E_D|$, Eq. (\ref{oddpart}) can be approximated by
\begin{equation}
\label{oddpart_approx}
G_{odd,22}(E)\simeq\frac{- \Gamma_1 \Gamma_2 \Gamma_3 \Delta(1-\frac{2 E^2}{\Delta \Gamma_1})}{E E_D^3},
\end{equation}
which gives the following expression for $\mathcal{D}_{22}$:
\begin{equation}
\label{d_large_ed}
\mathcal{D}_{22}\simeq\frac{- e \Gamma_1 \Gamma_2 \Gamma_3 \Delta }{2 h T E_D^3} \Big(1-\frac{2 \pi^2 k_B^2 T^2}{3 \Delta \Gamma_1}\Big).
\end{equation}
We can therefore distinguish two regimes according to whether the temperature is much smaller or much larger than $\sqrt{\frac{3 \Delta \Gamma_1}{2 k_B^2 \pi^2}}$. In the case where $(k_B T) \gg \sqrt{\frac{3 \Delta \Gamma_1}{2 \pi^2}}$ the sign of $\mathcal{D}_{22}$ is the same as $E_D$, otherwise it is the opposite.

It is interesting to analyze the behavior of the thermopower for systems similar to the one considered so far. Some remarks are in order. When one of the normal leads is removed to realize a two-terminal system, as shown in Appendix~\ref{2term}, the thermopower has the same sign in both $s$- and $p$-wave cases and the difference between the two cases is merely quantitative.
In particular, the thermopower decays in both cases with a power law as a function of $E_D$.
Similarly, as shown in Appendix~\ref{0enof}, no sign change occurs in the thermopower of a three-terminal setup where a $s$-wave superconducting wire hosts a fermionic zero-energy impurity end state (instead of an MBS).
This means that the thermopower of a three-terminal setup can distinguish between an MBS and an ordinary zero-energy fermionic state.
On the other hand, in Appendix~\ref{2lqd} we showed that if the QD has two levels, one of which is pinned at the Fermi energy, a sign change of the thermopower can occur even with an $s$-wave superconducting wire.
This shows that the sign change is not characteristic of the presence of an MBS.

%%%%%%%%%
\section{Tight-binding model}\label{tightbind}
\begin{figure}[tb]
\centering
\includegraphics[width=0.9\columnwidth, keepaspectratio]{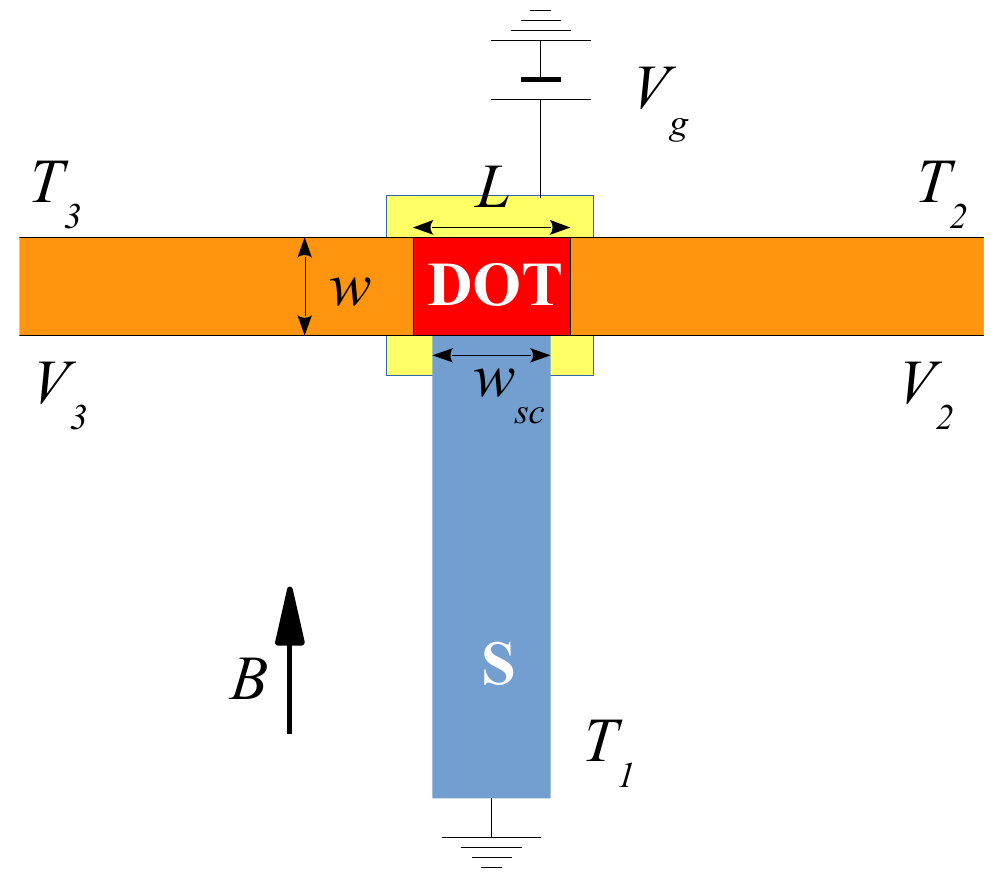}
\caption{Scheme of a T-junction formed by two semiconducting nanowires. The lower one (blue) is proximized by a superconductor and has finite spin-orbit interaction. An in-plane magnetic field perpendicular to the spin-orbit axis is applied to the whole system. Two barriers at distance $L$ form a multi-level quantum dot (red) in the upper nanowire (orange). The QD levels can be adjusted by a bottom gate (yellow) with applied voltage $V_g$.} 
\label{fig:real_setup}
\end{figure}

It is important to check to which extent the features found in the previous section survive in a more realistic situation. 
To this aim we consider a two-dimensional tight-binding spinful Hamiltonian that describes a T-junction of semiconducting nanowires\cite{valentini2014}. 
A schematic picture of this model is drawn in Fig. \ref{fig:real_setup}.
The lower (vertical) superconducting nanowire has a strong spin-orbit coupling, a Zeeman field orthogonal to its axis and a superconducting gap which is supposed to be induced by the proximity of an $s$-wave superconducting layer\cite{lutchyn2010,oreg2010}.
By varying the Zeeman field one can access the topological phase which is characterized by the presence of MBS.
In the upper (horizontal) normal nanowire, where only a Zeeman field is present, a (multi-level) QD is created by introducing two barriers.
A gate voltage is assumed to be present in the QD region, of length $L$, which induces an electrostatic potential that shifts the levels of the QD.
The superconducting nanowire is attached to the upper nanowire in between the two barriers to form the T-junction.
Far from the topological phase transition, we expect that the system is described by the simple 1D model introduced in Sec.~\ref{1d_model}. 

The Hamiltonian of the lower superconducting nanowire, of width $w$, reads
\begin{eqnarray}\label{eq:hamiltnw}
\hat {\cal H}_{\rm S} &=& -t\sum_{\langle i,j\rangle,\sigma}\hat c^\dag_{i,\sigma}\hat c_{j,\sigma} + (\varepsilon_0-\mu)\sum_{ i,\sigma}\hat 
c^\dag_{i,\sigma}\hat c_{i,\sigma} \nonumber\\
&& + i\lambda_R \sum_{\langle i,j\rangle,\sigma,\sigma'} (\nu'_{ij}\sigma^x_{\sigma\sigma'} - \nu_{ij}\sigma^y_{\sigma\sigma'})\hat 
c^\dag_{i,\sigma}\hat c_{j,\sigma'} \\
&&  + B \sum_{i,\sigma,\sigma'}\sigma^x_{\sigma\sigma'}\hat c^\dag_{i,\sigma}\hat c_{i,\sigma'} + \sum_{ i}\left[\Delta~\hat c^\dag_{i,\us}\hat c^\dag_{i,\ds} + \text{H.c.}\right]\;.\nonumber
\end{eqnarray}
Here $t$ is the hopping energy, $\varepsilon_0= 4 t$ is a uniform on-site energy, which sets the zero of energy,  $\lambda_R$ is the Rashba Spin-Orbit (SO) 
coupling strength, $B$ is the Zeeman field along the wire, $\Delta$ is the induced superconducting pairing, $\sigma^{i}$ are spin-1/2 Pauli matrices, $\nu_{i j} = \hat{\bm x} \cdot \hat {\bm d}_{i j}$, and 
$\nu'_{i j} = \hat{\bm y} \cdot \hat {\bm d}_{i j}$ with $\hat {\bm d}_{i j} = ({\bm r}_i - {\bm r}_j)/|{\bm r}_i - {\bm r}_j|$ being the unit vector connecting 
site $j$ to site $i$.

The upper normal nanowire is characterized by the same parameters as the lower one, but without SO and superconducting couplings, and its Hamiltonian reads
\begin{eqnarray}\label{eq:hamiltpr}
\hat {\cal H}_{\rm N} &=& -t\sum_{\langle i,j\rangle,\sigma}\hat c^\dag_{i,\sigma}\hat c_{j,\sigma} + (\varepsilon_0-\mu)\sum_{ i,\sigma}\hat 
c^\dag_{i,\sigma}\hat c_{i,\sigma} \\
&&  + B \sum_{i,\sigma,\sigma'}\sigma^x_{\sigma\sigma'}\hat c^\dag_{i,\sigma}\hat c_{i,\sigma'}
+V_g \sum_{ i\in {\rm QD},\sigma}\hat c^\dag_{i,\sigma}\hat c_{i,\sigma} ~, \nonumber
\end{eqnarray}
where the last term, only present in the QD region, changes the on-site energy by the additional quantity $V_g$, which represents the effect of a gate voltage.
The barriers defining the QD are accounted for by the following term: 
\begin{eqnarray}\label{eq:hamiltc}
\hat {\cal H}_{\rm b} &=& -t \left[ (\gamma_{L}-1) \sum_{\langle i,j\rangle,\sigma}^{({\rm b_L})}\hat c^\dag_{i,\sigma}\hat c_{j,\sigma} + 
(\gamma_{R}-1) \sum_{\langle i,j\rangle,\sigma}^{({\rm b_R})}\hat c^\dag_{i,\sigma}\hat c_{j,\sigma}  \right. \nonumber\\
&&\left. +\gamma_{D} \sum_{\langle i,j\rangle,\sigma}^{({\rm b_D})}\hat c^\dag_{i,\sigma}\hat c_{j,\sigma} \right] +
\text{H.c.}, 
\end{eqnarray}
where the superscripts $({\rm b_L})$, $({\rm b_R})$, and $({\rm b_D})$ in the symbol of the sum indicate that  the site $i,j$ are at the interfaces between the QD region and the rest of the nanowires.
The parameters $\gamma_{L}$, $\gamma_{R}$, and $\gamma_{D}$ are the strengths of the coupling to the left, right, and below of the QD, respectively. %
The complete Hamiltonian of the system then reads
$
\hat {\cal H} = \hat {\cal H}_{\rm S} + \hat{\cal H}_{\rm N} + \hat {\cal H}_{\rm b} ~.
$

\begin{figure}[tb]
\centering
\includegraphics[width=\columnwidth, keepaspectratio]{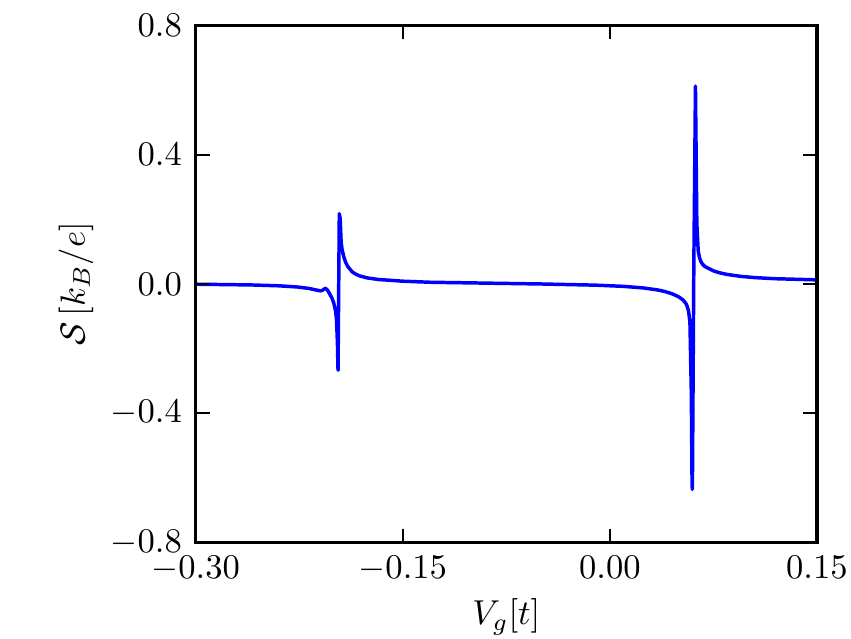}
\caption{Thermopower $\mathcal{S}$ of the realistic model as a function of $V_g$ for $B=0.08 t$ (trivial phase).}
\label{fig:triv}
\end{figure}

\begin{figure}[tb]
\centering
\includegraphics[width=\columnwidth, keepaspectratio]{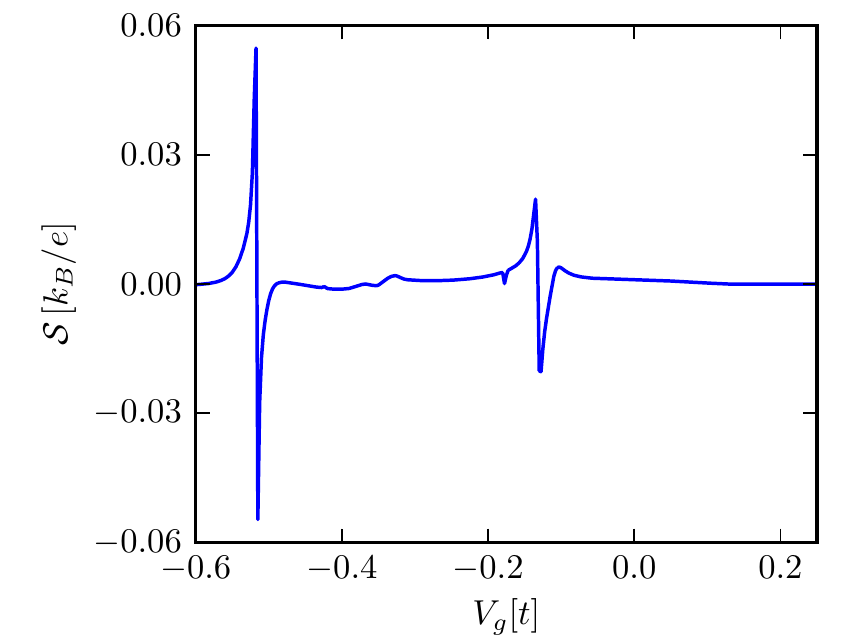}
\caption{Thermopower $\mathcal{S}$ of the realistic model as a function of $V_g$ for $B=0.14 t$ (topological phase).}
\label{fig:topo}
\end{figure}

We can now study the local thermopower defined in Eq. (\ref{eq::thermo}) as a function of the gate voltage $V_g$ for different values of the magnetic field.
We assume tunneling couplings to the left and right ($\gamma_L = 0.1$ and $\gamma_R = 0.05$), strong coupling with the superconducting nanowire ($\gamma_D = 1$) and a distance between barriers of $L=10$ sites.
Furthermore, we set both nanowire's width to be $w = 10$ sites, which implies single-channel transport for the parameters chosen here ($\mu=0.1 t$, $\lambda_R=0.1 t$ and $\Delta=0.1 t$). The phase transition occurs at $B\sim 0.10 t$.
As an example we plot a trivial case ($B=0.08 t$) in Fig. \ref{fig:triv} and a topological one ($B=0.14 t$) in Fig. \ref{fig:topo} for very small temperatures ($T=0.0001 t$).
In Fig.~\ref{fig:triv} we observe two regions around $V_g\sim -0.19 t$ and $V_g \sim 0.06 t$  where, for increasing $V_g$, the thermopower starts with a small negative value, drops rapidly to a minimum and then linearly grows, crosses zero to reach a maximum that is close to the opposite of the minimum and finally reaches a small positive value.
For both values of $V_g$ for which $\mathcal{S}$ crosses zero ($V_g\sim -0.19 t$ and $0.06 t$), we checked that the conductance $G_{22}(E)$ is symmetric with respect to zero.
This proves that each of these values of $V_g$ corresponds to having one of the levels of the (multi-level) QD aligned with the Fermi energy (this can be understood by looking at Fig.~\ref{fig:kernel} (upper panel) where $G_{22}(E)$ is not symmetric since the QD level $E_D\ne 0$, i. e., is not at the Fermi energy).
The curve in Fig. \ref{fig:topo} is slightly more structured, but we focus on two regions around $V_g\sim-0.52$ and $V_g\sim-0.13 t$ where the behavior is opposite with respect to the one just discussed, i. e., for increasing $V_g$ the thermopower starts with a small positive value, grows to a maximum and then linearly drops, crosses zero to reach a minimum that is close to the opposite of the maximum and finally reaches a small negative value.
This shows that the behavior of the thermopower when one of the QD levels crosses the Fermi energy is consistent with what we found in the previous section.
More precisely, the topological case is very close (and perfectly compatible in a certain range of $V_g$) to the $p$-wave case discussed before, while the trivial case is totally compatible with the $s$-wave case. 

\begin{figure}[tb]
\centering
\subfloat[][]{\label{fig:temp_triv}
\includegraphics[width=0.8\columnwidth, keepaspectratio]{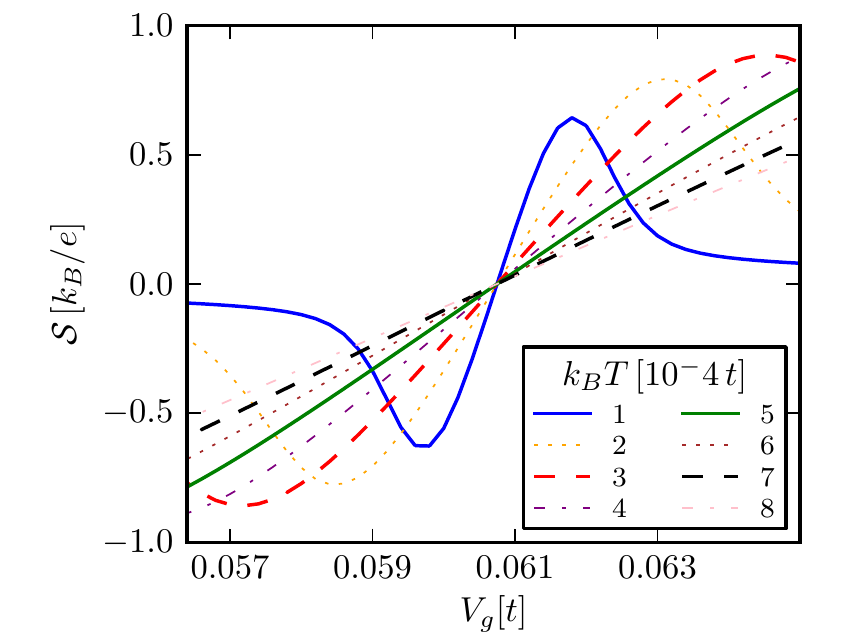}} \\
\subfloat[][]{\label{fig:temp_topo}
\includegraphics[width=0.8\columnwidth, keepaspectratio]{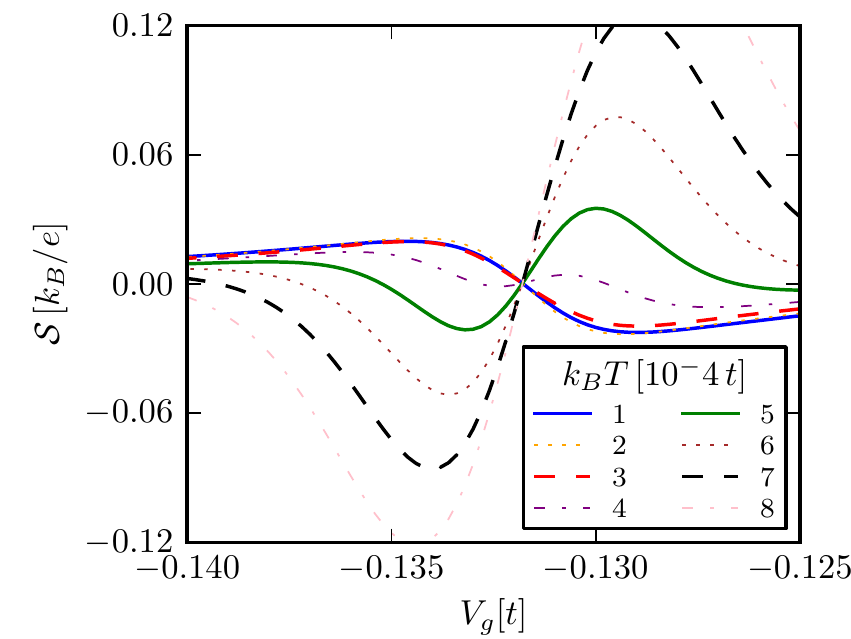}}
\caption{$\mathcal{S}$ as a function of $V_g$ for $B=0.08 t$ (\ref{fig:temp_triv}) and $B=0.14 t$ (\ref{fig:temp_topo}) for different temperatures.}
\label{fig:temp_all}
\end{figure}

We also explored the dependence of $\mathcal{S}$ on temperature in a smaller range of $V_g$ around a crossing with zero both for the trivial [Fig. \ref{fig:temp_triv}] and topological [Fig. \ref{fig:temp_topo}] phases. From Fig. \ref{fig:temp_triv} one can notice that for higher temperatures the thermopower keeps the same qualitatively behavior as in Fig. \ref{fig:triv}, even though its maximum gets bigger and is pushed away from the crossing with zero. 
On the contrary, for the topological case of Fig. \ref{fig:temp_topo}  the thermopower changes sign going to higher temperatures 
showing the same behavior as in the simple model of Sec. \ref{results}, including the intermediate behavior where $\mathcal{S}$ changes sign multiple times.
We can observe that the maximal thermopower is increased by rising the temperatures in all regimes. We also notice that the thermopower is no longer anti-symmetric with respect to the crossing and this feature appears more visible for higher temperatures. We attribute this behavior to the presence of multiple levels in the QD instead of the single-level QD of Sec. \ref{1d_model}, where $\mathcal{S}$ is an odd function of $E_D$. 
Assuming an induced gap of $250 \mu$eV, \cite{mourik2012} we estimate that the temperature $T= 10^{-4} t/{\rm k_B}$ corresponds to around $10$ mK.

\begin{figure}[tb]
\centering
\includegraphics[width=\columnwidth, keepaspectratio]{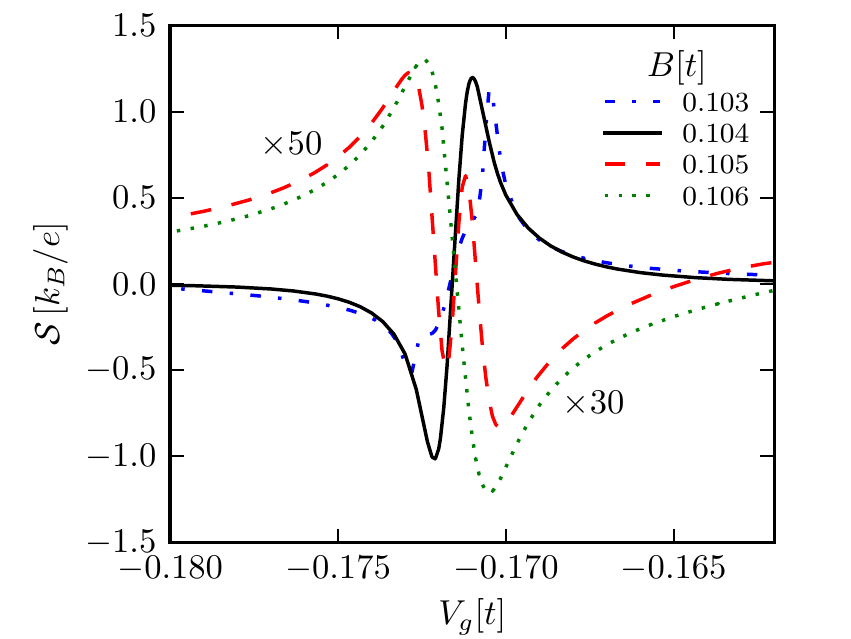} 
\caption{$\mathcal{S}$ as a function of $V_g$ for different values of $B$ near the phase transition. The curves are shifted horizontally to cross zero at the same point as for $B=0.103 t$.}
\label{fig:transiz}
\end{figure}

Finally, we study the thermopower as a function of $V_g$ for different values of the Zeeman field $B$ through the topological phase transition. First of all, we notice that the crossings with $\mathcal{S}=0$ shift linearly with $B$, meaning that only one spin species is involved in each level. Far from the phase transition (which occurs at $B_{tr} \sim 0.1039978 t$) the shape of the curves does not change for all magnetic fields in each phase, i.e., for $B\ll B_{tr}$ the curves are similar to the one in Fig. \ref{fig:triv}, while for $B\gg B_{tr}$ the curves are similar to the one in Fig. \ref{fig:topo}. In Fig. \ref{fig:transiz}, we show what happens closer to the transition, in particular we plot the curves for $B=0.103 t$ (trivial phase) and for $B=0.104, 0.105, 0.106 t$ (topological phase) shifted horizontally in order for all the curves to cross $\mathcal{S}=0$ at the same point.
We observe that the curve relative to $B=0.104 t$ has positive slope at the crossing point as for the high temperature case discussed above, while the curve for $B=0.105 t$ is more structured as in the intermediate temperature case and the curve for $B=0.106 t$ has the usual low temperature, topological behavior. The behavior at $B=0.104 t$ and $B=0.105 t$ can be explained by noticing that near the phase transition the band gap shrinks and it becomes comparable to or smaller than $\sqrt{\frac{2 \pi^2 (k_B T)^2}{3 \Gamma_1}}$ [see the discussion of Eq. \ref{d_large_ed}]. 
This actually implies that the onset of the intermediate regime can be used as a signature of the phase transition which allows an approximate determination of $B_{tr}$. Indeed, referring to Fig. \ref{fig:transiz}, by exploring the behavior of $\mathcal{S}$ while varying $B$ one can be sure that the transition has already occurred at $B=0.105 t$, giving an accuracy of around $1\%$ of $B_{tr}$ with the parameters we used.

\section{Conclusions}
In summary, we have studied the thermopower of a three-terminal junction composed of a quantum dot attached to two normal leads and a topological superconducting wire. We analyzed a simple 1D model where the superconductor could be either $p$-wave or $s$-wave focusing on the thermopower as a function of the QD level $E_D$. We showed that for small temperatures the thermopower in the $p$-wave case has opposite sign with respect to the $s$-wave case and that this behavior is strongly dependent on the temperature.
Also, for intermediate temperatures more complex structures appear.
We explain this low-temperature behavior with the presence of a Majorana bound state in the $p$-wave case and we identify the energy scale that controls the sign of the thermopower with  $\sqrt{\frac{3 \Delta \Gamma_1}{2 \pi^2}}$, when the energy level of the QD is well off resonance.
This allows us to distinguish two different regimes according to whether $k_B T$ is much bigger or smaller than this scale.
Furthermore, we confirm this behavior in a more realistic 2D tight-binding model studying the thermopower as a function of a gate voltage controlling the multi-level quantum dot. In the topological phase the thermopower behaves as in the $p$-wave case of the simple model, while the trivial phase resembles the $s$-wave case. The dependence on the temperature is also confirmed. We also study the thermopower across the topological phase transition, concluding that the measurement of such quantity
can give an approximate value of the magnetic field at which the transition occurs.

Finally, we checked that this behavior of the thermopower is not reproducible by a similar system with a zero-energy normal (fermionic) resonance between the QD and the superconductor in the $s$-wave case, but it is qualitatively similar to that of an ``exotic" two-level quantum dot where one of them is kept fixed at the Fermi energy while the other is moving, in the $s$-wave case. From this, we can conclude that these features are not peculiar to the presence of a Majorana bound state.
We found, in addition, that when one of the normal leads is removed (realizing a two-terminal setup) the thermopower decays with a power law both for the $p$- and the $s$-wave case.
We conclude by noting that our approach allows to go beyond the linear-response regime, for example for studying the effects of a large temperature bias.

\begin{appendix}
\section{Two-terminal setup}\label{2term}
For completeness, in this appendix, we study the thermopower of a simpler two-terminal setup in which the QD attached to the (topological-)superconducting wire is coupled to only one normal lead.
This system has been already studied in Ref.~\onlinecite{leijnse2014} using a different approach.
Here we use the same model as in Sec.~\ref{1d_model}, keeping in mind that the scattering matrix, Eq. (\ref{qdmatrix}), is for two terminals so that one can only define a local thermopower, namely, 
\begin{equation} \label{eq::thermo_2term}
 \left.\mathcal{S}=-\frac{\Delta V_2}{\Delta T_2} \right|_{J_N^1=0,J_N^2=0}=\frac{\mathcal{D}_{22}}{\mathcal{G}_{22}}.
\end{equation}
Using the definition (\ref{D}), one can notice that the subgap contributions to the integral defining $\mathcal{D}_{22}$ are zero due to the intrinsic particle-hole symmetry of  the Bogoliubov-de Gennes approach and the unitarity of the scattering matrix. On the contrary, these contributions are relevant to the integral defining $\mathcal{G}_{22}$ [Eq.~(\ref{G})]. The above-gap contributions are qualitatively the same in the $s$-wave and $p$-wave cases because the MBS influences mainly the subgap conductance.
This means that no sign-inversion of the thermopower is possible even when the parameters are varied.  The big difference between the $s$-wave and the $p$-wave cases is not qualitative, but quantitative.
Indeed, the maximum of the thermopower as a function of the dot level is much higher in the $s$-wave case essentially because the electrical conductance is smaller due to the absence of the Majorana zero-bias peak that dominates $\mathcal{G}_{22}$.
\begin{figure}[tb]
\centering
\includegraphics[width=0.9\columnwidth, keepaspectratio]{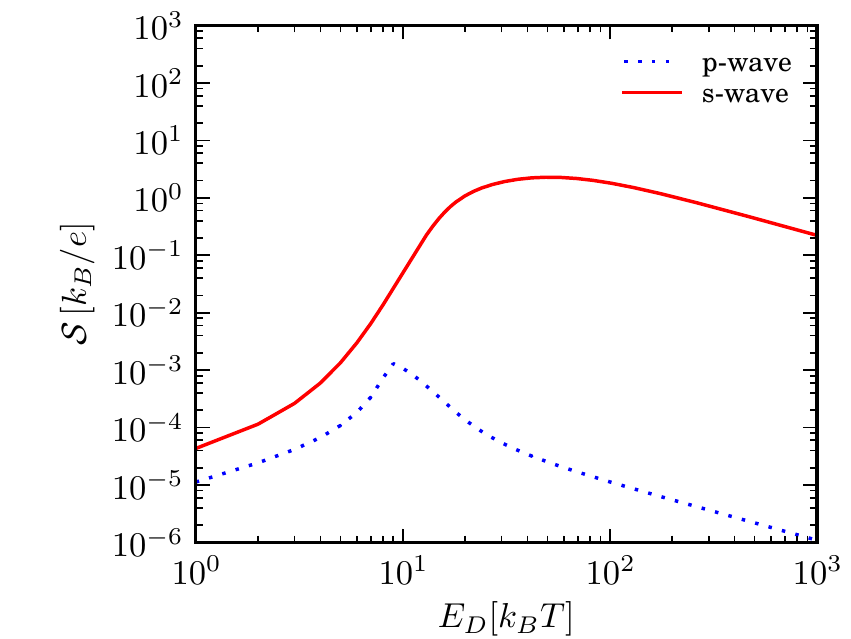}
\caption{$\mathcal{S}$ as a function of $E_D$ for a two-terminal setup, with $\Delta=10 k_B T$, $\Gamma_1= 10 k_B T$, $\Gamma_2=0.1 k_B T$. Red (blue) solid (dotted) line refers to an s-wave (p-wave) superconducting wire.} 
\label{fig:2term}
\end{figure}
In Fig. \ref{fig:2term}, we show a bilogarithmic plot of the thermopower as a function of the QD level both in the $s$-wave and the $p$-wave cases. This makes clear that, in both cases, the thermopower decays with a power law as a function of $E_D$.

\section{Zero-energy ordinary-fermionic resonance}
\label{0enof}
\begin{figure}[tb]
\centering
\includegraphics[width=0.9\columnwidth, keepaspectratio]{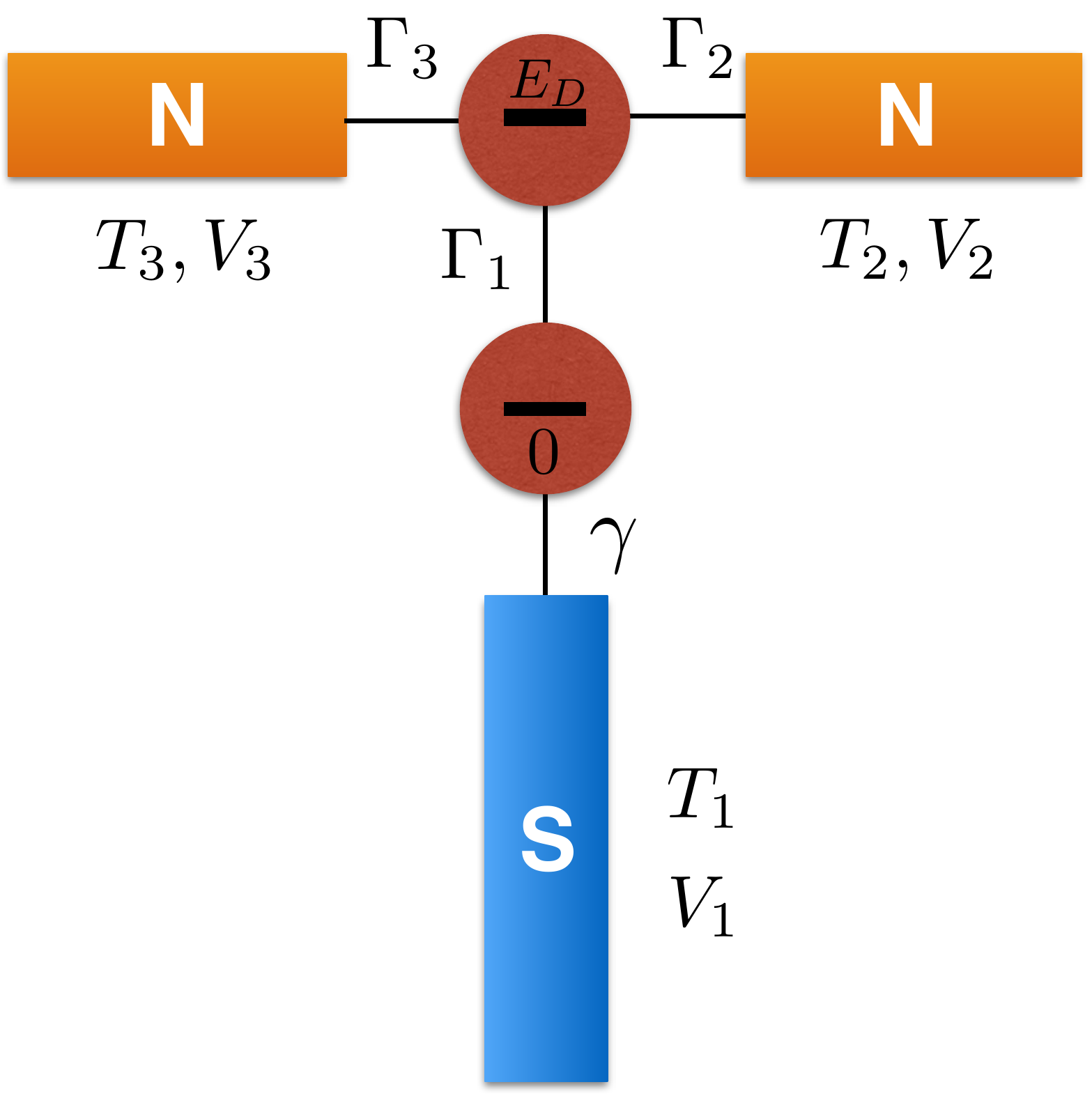}
\caption{Scheme of the zero-energy impurity end state setup. The upper QD coupled to two normal leads (N) is now connected to a trivial superconductor (S) through an additional zero-energy-level QD.}
\label{ordinary_fig}
\end{figure}

\begin{figure}[tb]
\centering
\includegraphics[width=0.9\columnwidth, keepaspectratio]{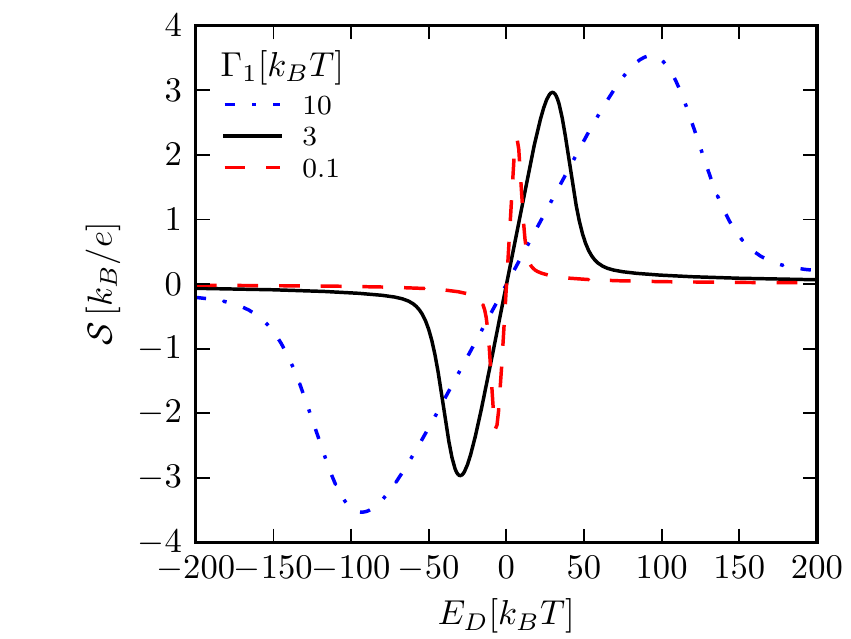}
\caption{$\mathcal{S}$ as a function of $E_D$ for the system sketched in Fig.~\ref{ordinary_fig}, with $\Delta=100 k_B T$, $\Gamma_2=0.1 k_B T$, $\Gamma_3=0.1 k_B T$, $\gamma=0.1 k_B T$, and different values of $\Gamma_1$.} 
\label{ordinary_thermo}
\end{figure}

In this appendix, we show that the behavior of the thermopower reported in Sec.~\ref{results} for a MBS cannot be reproduced by a trivial superconducting wire which hosts a fermionic zero-energy impurity end state.
In particular, we study the thermopower of the system sketched in Fig.~\ref{ordinary_fig} where the topological superconducting wire of Fig.~\ref{fig:setup_simple} is replaced by an $s$-wave superconducting wire coupled to a single-level QD, whose energy position is fixed to zero.
Here we use the model detailed in Sec.~\ref{1d_model} where now the scattering matrix describing the double-QD system is obtained by composing the matrices relative to the two QDs, the lower one comprising two couplings (of equal strength $\gamma$) and the upper comprising three couplings $\Gamma_1$, $\Gamma_2$, and $\Gamma_3$.
In Fig.~\ref{ordinary_thermo}, the thermopower $\mathcal{S}$ defined as in Sec.~\ref{1d_model} is plotted as a function of the upper-QD level $E_D$ for various values of the coupling $\Gamma_1$. Although quantitatively different, the three curves are qualitatively similar and in particular no sign-change occurs by varying $\Gamma_1$, in contrast to what happens in the presence of the MBS (see Fig.~\ref{fig:g1} for comparison).

\section{Two-level Quantum Dot}
\label{2lqd}
\begin{figure}[tb]
\centering
\includegraphics[width=0.9\columnwidth, keepaspectratio]{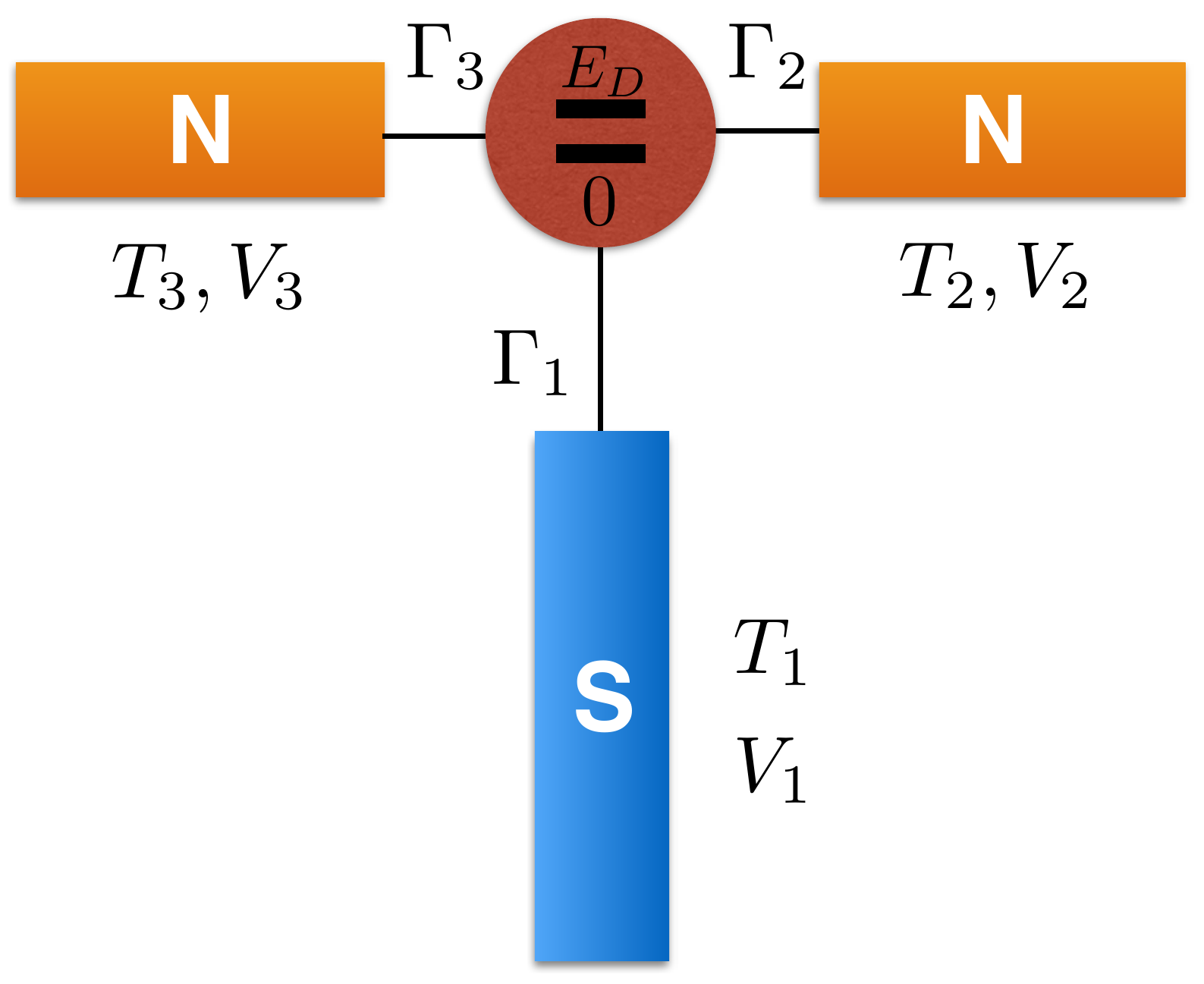}
\caption{Scheme of the two-level QD setup. The QD, with one level at energy $E_D$ and the other at zero-energy, is coupled to two normal leads (N) and a trivial superconducting wire (S).}
\label{fig:setup_double}
\end{figure}

In this appendix, we show that the behavior of the thermopower of the three-terminal setup with the topological superconductor described in Sec. \ref{results} is not peculiar to such a system. Here we prove that it is qualitatively the same as for the system composed of a two-level QD (where one level is pinned at the Fermi energy) attached to two normal leads
and an $s$-wave superconductor, as shown in Fig. \ref{fig:setup_double}. 
For simplicity, in our calculations we choose the same couplings $\Gamma_i$ between the two levels and each lead. Following the same procedure and adopting the same definitions as in Sec. \ref{1d_model}, one can show that the behavior of the thermopower as a function of the not-pinned level $E_D$ is the same as for the $p$-wave, single-level case. 
\begin{figure}[tb]
\centering
\includegraphics[width=0.9\columnwidth, keepaspectratio]{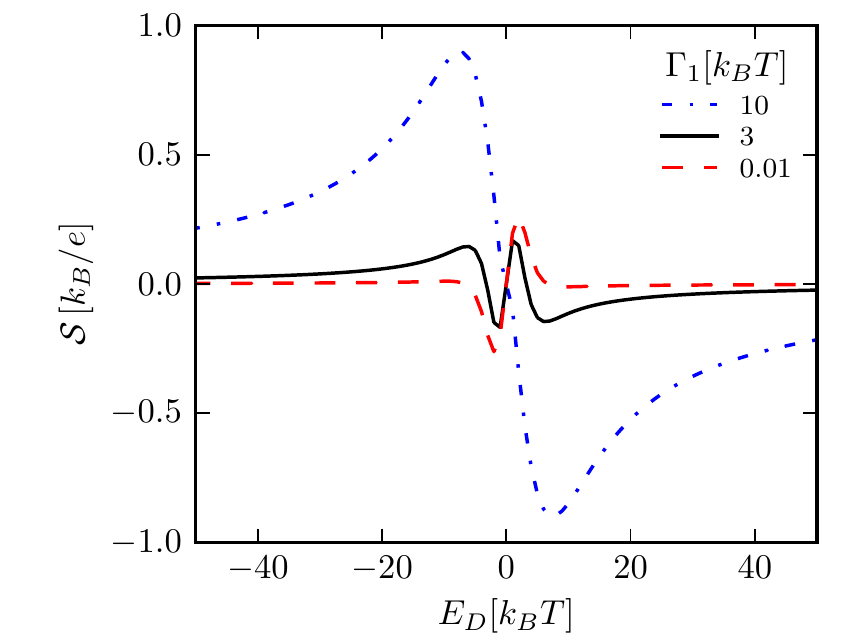}
\caption{$\mathcal{S}$ as a function of $E_D$ for the double-level QD sketched in Fig.~\ref{fig:setup_double}, with $\Delta=100 k_B T$, $\Gamma_2=0.1 k_B T$, $\Gamma_3=0.1 k_B T$, and different values of $\Gamma_1$ .} 
\label{fig:double}
\end{figure}
As an example, in Fig. \ref{fig:double}, we plot the thermopower $\mathcal{S}$ as a function of $E_D$ for different values of the coupling to the superconductor $\Gamma_1$ and we notice that the sign inversion and the intermediate regime are present also in this topologically trivial setup.
Note, however, that $E_D$ in practice is changed by varying a gate voltage which would act on both levels of the QD. Therefore the situation described above whereby one level only is changed, while the other is pinned at the Fermi energy, is unrealistic.

\end{appendix}

%%%%%%%
\section*{Acknowledgements}
We would like to acknowledge fruitful discussions with Y. Asano.
This work has been supported by EU project ThermiQ, by MIUR-PRIN: "Collective quantum phenomena: from strongly correlated systems to quantum simulators", by the EU project COST Action MP1209 "Thermodynamics in the quantum regime", and by the EU project COST action MP1201 "NanoSC".

\end{document}